**Quantum liquid from strange frustration in the trimer magnet Ba$_4$Ir$_3$O$_{10}$**

Gang Cao[1*], Hao Zheng[1], Hengdi Zhao[1], Yifei Ni[1], Christopher. A. Pocs[1], Yu Zhang[1], Feng Ye[2], Christina Hoffmann[2], Xiaoping Wang[2], Minhyea Lee[1], Michael Hermele[1,3] and Itamar Kimchi[1,3,4*]

[1]Department of Physics, University of Colorado at Boulder, Boulder, CO 80309, USA

[2] Neutron Scattering Division, Oak Ridge National Laboratory, Oak Ridge, TN 37831, USA

[3] Center for Theory of Quantum Matter, University of Colorado, Boulder, CO 80309, USA

[4] JILA, NIST and Department of Physics, University of Colorado, Boulder, CO 80309, USA

**Abstract.** Quantum spin systems such as magnetic insulators usually show magnetic order, but such classical states can give way to *quantum liquids with exotic entanglement* through two known mechanisms of frustration: geometric frustration in lattices with triangle motifs, and spin-orbit-coupling frustration in the exactly solvable quantum liquid of Kitaev's honeycomb lattice. Here we present the experimental observation of a new kind of frustrated quantum liquid arising in an unlikely place: the magnetic insulator $Ba_4Ir_3O_{10}$ where $Ir_3O_{12}$ trimers form an unfrustrated square lattice. The crystal structure shows no apparent spin chains. Experimentally we find a quantum liquid state persisting down to 0.2 K that is stabilized by strong antiferromagnetic interaction with Curie-Weiss temperature ranging from -766 K to -169 K due to magnetic anisotropy. The anisotropy-averaged frustration parameter is 2000, seldom seen in iridates. Heat capacity and thermal conductivity are both linear at low temperatures, a familiar feature in metals but here in an insulator pointing to an exotic quantum liquid state; a mere 2% Sr substitution for Ba produces long-range order at 130 K and destroys the linear-T features. Although the $Ir^{4+}(5d^5)$ ions in $Ba_4Ir_3O_{10}$ appear to form $Ir_3O_{12}$ trimers of face-sharing $IrO_6$ octahedra, we propose that intra-trimer exchange is reduced and the lattice recombines into an array of coupled 1D chains with additional spins. An extreme limit of decoupled 1D chains can explain most but not all of the striking experimental observations, indicating that the inter-chain coupling plays an important role in the frustration mechanism leading to this quantum liquid.

* Corresponding authors; gang.cao@colorado.edu and itamar.kimchi@colorado.edu

**Introduction.** Quantum spin systems can enter unusual quantum phases of matter known as quantum liquids. The first such quantum liquids were discovered in one-dimensional systems that are known as Tomonaga-Luttinger liquids; a second class of quantum liquids can occur in two or higher dimensions and goes by the name *quantum spin liquids* [1]. These are quantum phases with fractionalized excitations that cannot be adiabatically connected to a stack of 1D systems. The usual fate of 2D and 3D magnets is magnetic order, so to enter a quantum fluctuating liquid phase the competing magnetic orders must all be energetically avoided through *magnetic frustration*. Frustration as a mechanism for quantum liquids was first discussed by Anderson [2]. One key observation is that since spins often prefer to anti-align with their neighbors, lattices that contain triangles give an energetic degeneracy for anti-aligned spins. Indeed, spin liquid candidates have been found on the triangular lattice e.g. in organic compounds [1,3] and on the related kagome lattice e.g. in herbertsmithite [1,3]. A more recent development was the theoretical discovery by Kitaev [4] of a second kind of frustration mechanism, manifested in an exactly solvable spin liquid model on the honeycomb lattice with strong spin-orbit interactions (SOI). SOI is especially significant in heavy magnetic elements such as correlated oxides with $Ir^{4+}$ ($5d^5$) ions, and the search for Kitaev's spin liquid in honeycomb iridates is ongoing [5-9]. No other magnetic frustration mechanisms for quantum liquids in magnetic Mott insulators have so far been established.

Here we report an exotic extraordinarily frustrated spin state occurring in an unlikely place – $Ba_4Ir_3O_{10}$ single crystals where $Ir_3O_{12}$ trimers form an unfrustrated 2D square lattice. It is emphasized that the crystal structure provides no obvious settings for the formation of spin chains (**Fig. 1**). This insulating iridate with tetravalent $Ir^{4+}$ ($5d^5$) ions presents concurrently an absence of long-range magnetic order above 0.2 K and extraordinarily strong antiferromagnetic (AFM) exchange interactions characterized by Curie-Weiss temperatures $\theta_{CW}$ of (-766 K, -390 K, -169 K) along the (*a, b, c*) axes, respectively (**Fig. 2 a, b**). The contrasting values of $\theta_{CW}$ and the possible Néel temperature $T_N = 0.2$ K result in an astonishing frustration parameter $f = 2000$, which is an anisotropy-averaged $f = |\theta_{CW}|/T_N = 3800, 1980$ and 845, indicating a quantum liquid state. For comparison a magnet with $f > 10$ is usually considered frustrated [10], and synthetic herbersmithite, an archetypical quantum spin liquid candidate [11], shows $f \geq 3600$. The quantum liquid state in $Ba_4Ir_3O_{10}$ is corroborated by a sizable, linear heat capacity at very low temperatures, suggesting that the absence of magnetic ordering is accompanied by gapless quantum-fluctuating excitations. The T-linear heat capacity extrapolates to a T=0 nonzero offset and associated extensive entropy, implying an emergent ultralow energy scale. Linear thermal conductivity is also observed at low temperatures, again consistent with the quantum liquid state. Finally, aiming to test the stability of this low energy state to lattice changes, we find that a mere 2% Sr substitution for Ba (($Ba_{0.98}Sr_{0.02})_4Ir_3O_{10}$) readily removes frustration and precipitates AFM order at 130 K accompanied by drastic changes in magnetization, heat capacity, and thermal conductivity, implying an extraordinary sensitivity of the quantum-fluctuating state in $Ba_4Ir_3O_{10}$ to the lattice, expected for immense frustration driven by the SOI together with the crystal structure. We propose a theoretical scenario of a quantum liquid state that emerges from a recombination of the $Ir_3O_{12}$ trimers and discuss it from the starting point of an array of one-dimensional Tomonaga-Luttinger Liquids coupled to additional spins. Its consistency with the striking experimental observations suggests that the quantum liquid in $Ba_4Ir_3O_{10}$ arises from a frustration mechanism, distinct from the usual geometric and spin-orbit (Kitaev honeycomb model) frustration mechanisms.

Iridates with $Ir^{4+}$ ($5d^5$) ions, in which the odd electron count produces a Kramer's doublet $J_{eff} = 1/2$ local moment, have been extensively discussed as fertile ground for SOI-driven frustrated quantum magnetism [1,3,5-9]. However empirically they typically show long-range magnetic order at temperatures comparable to the magnetic exchange energies, for instance 5 - 40 K in a group of the Kitaev quantum spin liquid candidates [1,3,6-9], indicating that the most dramatic effects of quantum fluctuations have not yet materialized (at least without further tuning [12]). Indeed, with one possible exception of the newly synthesized $H_3LiIr_2O_6$ [13,14], no other iridates with exchange energies of hundreds of Kelvin have been found to be paramagnetically quantum-fluctuating down through sub-Kelvin temperatures [1,3], until now.

**Results.** $Ba_4Ir_3O_{10}$ adopts a monoclinic structure with a $P2_1/c$ space group, consistent with some previous structural data [15] (but not all [16]) (See Supplementary Fig.1 in Supplementary Material (SM) for details). The crystal structure consists of $Ir_3O_{12}$ trimers of face-sharing $IrO_6$ octahedra that are vertex-linked to other trimers, forming wavelike 2D sheets in the *bc* plane that are stacked along the *a* axis with no connectivity between the sheets (**Figs.1a** and **1b**). The nearly identical Ir-O bond distances for both Ir1 and Ir2 sites within each trimer implies the same valence state for Ir1 and Ir2 ions in the trimers. Electrical resistivity shows a clear insulating state across the entire temperature range measured up to 400 K (see Supplementary Fig.5 in SM).

As shown in **Fig.2a**, the magnetic susceptibility, $\chi$, of $Ba_4Ir_3O_{10}$ for all three principal crystalline axes exhibits no anomalies or sign of magnetic order down to 1.7 K.

No hysteresis behavior is seen in $\chi$. The paramagnetic behavior perfectly follows the Curie-Weiss law for a temperature range of 100 K – 350 K, which is illustrated in a plot of $1/\Delta\chi$. (**Fig.2b**). Here $\Delta\chi = \chi - \chi_o$, with $\chi_o$ the temperature-independent susceptibility, expected to arise in this d-electron system as the Van Vleck susceptibility from high energy crystal field levels of $Ir^{4+}$ (with energies of order eV, much higher than the highest measurement temperatures). The Curie-Weiss temperature $\theta_{CW}$ (intercept on the horizontal axis) is determined to be -766 K, -390 K and -169 K for the *a*, *b* and *c* axis, respectively. These extracted values (which vary due to the anisotropic exchange interactions) are comparable to the temperatures measured, implying that the susceptibility is not in its high temperature asymptotic form, hence $\theta_{CW}$ should be considered as giving an order of magnitude for the exchanges. The unusually large magnitudes of $\theta_{CW}$ reveal an extraordinarily strong tendency for an AFM order – but, no long-range magnetic order is discerned above 0.2 K, according to the heat capacity data discussed below. These results reveal an exceptionally large anisotropy-averaged frustration parameter $f$ = 2000 [=(3800+1980+840)/3], suggesting an unusually robust, large spin-liquid regime. Moreover, the effective moment, $\mu_{eff}$, estimated from the Curie-Weiss fit is 1.78, 0.80 and 0.4 $\mu_B$/Ir for the *a*, *b* and *c* axis, respectively (c.f. $\mu_{eff}$ = 1.73 $\mu_B$ for an isolated $S$ = ½ moment with no SOI). These values are comparable to or greater than those of magnetically ordered iridates such as 0.13 $\mu_B$/Ir for $BaIrO_3$ [17] and 0.50 $\mu_B$/Ir for $Sr_2IrO_4$ [18], supporting the expected picture of a SOI $S_{eff}$ = ½ local moment per Ir site.

The heat capacity, C(T), is thus examined focusing on low temperatures (0.05 – 4 K). No sign of an ordering transition is found above T* = 0.2 K (**Fig.3a**). Below 0.2 K, C(T) rises abruptly, indicating magnetic order. Schottky effects are ruled out by the anisotropic response of C(T) to an applied magnetic field, H: the upturn in C(T) below T* changes only slightly for H || a or c axis but drops abruptly at a critical field $\mu_o H = 0.20$ T for H || b axis (**Fig.3b**). Specifically, with increasing H (||b axis), C(T) decreases initially for $\mu_o H \leq 0.19$ T, and then precipitates a rapid downturn at a critical field $\mu_o H_c = 0.2$ T (Inset in **Fig. 3b**). The entropy removal below 4 K is estimated to be around 0.15 J/mole K at $\mu_o H_c = 0.2$ T, which suggests that $Ba_4Ir_3O_{10}$ behaves like a Fermi liquid metal where most of the entropy removal happens near a Fermi temperature, $T_F$, and the T-linear C(T) occurs at $T \ll T_F$.

We now focus on the low energy physics in the correlated paramagnet regime 0.2K $< T \ll \theta_{CW}$. The nature of the upturn below 0.2 K will be addressed elsewhere. C(T) presents a pronounced linear temperature dependence over a one-decade temperature span between 0.2 K and 2.5 K (**Fig.3a**). The linear slope is $\gamma = 17$ mJ/mole K$^2$. The T-linear C(T) extrapolates to a constant offset at T=0. Applying small magnetic fields up to a few tesla (**Fig.3c**) preserves the constant offset, and rounds off the lowest temperature portion of the T-linear C(T) only to an extent consistent with an increase in the magnetic order instability T* from 0.2 K to ~1 K due to a lowering of the magnetic order energy in the applied field. (Antiferromagnetic order should strongly couple to applied fields since low crystal symmetry generically gives magnetic order a nonzero net ferromagnetic moment.) This behavior, which is not at all expected for any conventional insulator, reveals an existence of large residual entropy despite such low temperatures and is consistent with a quantum liquid state (discussed below). Linear heat capacity is seen in some spin-½ spin-liquid candidates [e.g., 19], suggesting gapless excitations [20, 21]. Furthermore, the Wilson ratio expressed as

$$R_W \equiv 3\pi^2 k_B^2 \chi / \mu_B^2 \gamma \quad (1)$$

is estimated to be 43, far beyond the values (e.g., 1-6) typical of exchange-enhanced paramagnets, manifesting strong SOI. Interestingly, it is close to that of the hyperkagome $Na_4Ir_3O_8$ [22]; in fact, $R_W$ is significantly larger than 1 for almost all spin-liquid candidates [23]. In addition, the thermal conductivity also features a clear linear temperature dependence below 8 K, which disappears upon slight Sr substitution (discussed below). All these results point to an exotic quantum liquid state, with itinerant gapless excitations, arising at low energies from the effective SOI spin-1/2 moments in $Ba_4Ir_3O_{10}$.

This liquid state is strikingly sensitive to even slight lattice alterations, as evidenced in $(Ba_{1-x}Sr_x)_4Ir_3O_{10}$ where x = 0.01 and 0.02 corresponding to 1 and 2% Sr substitution for Ba. This nonmagnetic, isovalent substitution readily lifts the enormous frustration at x=0.01 and stabilizes a transition at $T_N$ = 130 K at x=0.02 revealed in the magnetization M(T) and heat capacity C(T) (**Fig.4a**) (Note that heat capacity is an effective measure of a bulk effect; the observed anomaly at $T_N$=130K in C(T) undoubtedly confirms an intrinsic phase transition). It is emphasized that the substitution only alters the lattice parameters but causes no further structural transition [see Supplementary Fig.1 in SM]. C(T) shows a kink at $T_N$ but no discernable anomaly at the lower temperature $T_M$, where M(T) exhibits a far more pronounced peak. A sharp metamagnetic transition is seen at $\mu_o H$ = 4.5 T along the *a*-axis below $T_M$ (Inset in **Fig.4a**). The sharp angular dependence of M for the *ab* plane confirms an easy axis primarily aligned along the *a* axis ($\perp$ 2D sheets) and a strong magnetic anisotropy in x = 0.02 below $T_M$ (**Fig.4b**). However, the saturation moment is very small, no more than 0.03 $\mu_B$/Ir on average. The *a*-axis Curie-Weiss temperature $\theta_{CW}$ is reduced to -130 K from -766 K, thus yielding a frustration parameter *f*=1, drastically reduced from 3800 or by over three orders of magnitude (**Fig.4c**). Also see Supplementary Fig.7

The low-temperature C(T) for x = 0.01 and 0.02 is no longer linear (**Fig.5a**), with a complete removal of residual entropy below 0.05 K and the upturn below 0.2 K. The thermal conductivity along the c-axis $k_c$ exhibits the pronounced linear dependence below 8 K for x = 0 and a distinct non-linear feature for x = 0.02 (**Fig.5b**). The behavior of $k_c$ is vastly different from that of other systems [24, 25], and is dominated by phonons that are strongly scattered by fluctuating spins. Indeed, $k_c$ for x = 0 exhibits no magnetic field dependence [see Supplemental Material for more data]. The disappearance of the low-temperature linear dependence in $k_c$(T) (together with C(T)) upon Sr doping implies the T-linearity should be associated with the quantum liquid behavior. Moreover, $k_c$(T) for x = 0.02 shows an anomaly at $T_N$ attributable to spin-phonon scattering, followed by a peak at 25 K; the absence of such a peak in the original $Ba_4Ir_3O_{10}$ independently implies the existence of strong spin fluctuations associated with the quantum liquid state (also see Supplementary Fig.6 in SM).

**Discussion.** We now turn to the low energy theoretical description. Usually the face-sharing geometry for $IrO_6$ octahedra is believed to produce stronger magnetic exchange, e.g. due to the shorter Ir-Ir distance (2.576 Å) [26]. Under this assumption the $Ir_3O_{12}$ trimers in $Ba_4Ir_3O_{10}$ would form the basic magnetic unit, with an effective spin-1/2 Kramer's doublet on each trimer, and the lattice of trimers would determine all low-energy properties. From this point of view, the magnetic phenomenology in $Ba_4Ir_3O_{10}$ is puzzling: its observed frustration is inconsistent with the trimer picture. To see the inconsistency, let us focus on the fact that trimers in $Ba_4Ir_3O_{10}$ are arranged in a diagonally distorted square lattice (**Fig.1**).

Neighboring trimers are coupled via corner-sharing IrO$_6$ octahedra, which are known to produce dominant Heisenberg antiferromagnetic exchange interactions (with small perturbations through distortions and Hund's coupling) [5]. Consequently, the Heisenberg antiferromagnet on this lattice would produce a robust Neel antiferromagnetic order, and with even small SOI or interlayer couplings, gives a high ordering temperature with much reduced frustration parameter. Clearly, this trimer picture cannot explain the observed $f$ = 3800. This point is further supported by the rapid loss of frustration upon the 2% Sr substitution in Ba$_4$Ir$_3$O$_{10}$ and the well-established, long-range magnetic order at T$_N$ = 183 K in the sister compound BaIrO$_3$ where analogous trimers couple into a distorted cubic lattice [17]. In both cases the observed T$_N$ is thousand times higher than T$_N$ in Ba$_4$Ir$_3$O$_{10}$ (also see Supplementary Figs.8 and 9 in SM).

The inability to explain the exotic frustration in Ba$_4$Ir$_3$O$_{10}$ via trimer-based pictures implies that a different theoretical starting point is needed and, in particular, involves some reduction, compared to conventional folklore, of the effective face-sharing intra-trimer exchange. The octahedral face-sharing geometry has not been as theoretically well explored with few recent exceptions [26, 27]. Two theoretical observations [26] are relevant to the local moment magnetism in Ba$_4$Ir$_3$O$_{10}$: **(1)** Ir$^{4+}$ face-sharing octahedra with either strong SOI or strong trigonal distortions produce spin exchange with an emergent spin-rotation symmetry. In Ba$_4$Ir$_3$O$_{10}$ the local crystal field can be approximated by purely trigonal stretching, so intra-trimer exchange is approximately captured by a single parameter, namely the coefficient of a Heisenberg term. **(2)** The face-sharing electron hopping integral *changes sign* as a function of octahedral distortions **(Fig. 6a)**. Tuning through crystal structures near the sign change gives rise to an unexpectedly weak Heisenberg intra-trimer exchange, which thus need not be larger than inter-trimer exchange.

Interestingly, this mechanism also automatically accounts for the high sensitivity to local crystalline distortions by the slight Sr substitution -- even small local distortions can drastically change the exchange through face sharing octahedra and thus change the effective lattice connectivity, allowing for local magnetic domains.

Now consider the artificial opposite limit with face-sharing intra-trimer exchange smaller than the corner-sharing inter-trimer exchange. Two thirds of the sites (trimer endpoints) arrange into an array of 1D zigzag chains oriented along the *c* axis (**Fig.6b**). The coupling within each chain occurs through corner-sharing octahedra, producing exchanges that should be well approximated by Heisenberg AFM exchange with planar anisotropy [5]. Within the broad energy regime where such an effective U(1) spin rotation symmetry appears, and when the sign of the anisotropy is easy-plane, each spin chain is described at low energy by the 1D Tomonaga-Luttinger theory of a highly fluctuating quantum liquid. This limit is consistent with some of striking experimental observations, in particular, a quantum liquid that is robust against magnetic order, and features T-linear $C(T)$ and $k$(T). (This scenario can also explain the difference between $Ba_4Ir_3O_{10}$ and $BaIrO_3$: each 1D chain in $Ba_4Ir_3O_{10}$ is replaced by a 2D lattice in $BaIrO_3$, leading to conventional magnetic order [16,17]) The remaining one third (trimer midpoint) sites are only weakly coupled in this limit; fully decoupled spins would contribute to magnetic susceptibility, which is consistent with the data, but more significantly to low-temperature heat capacity, which would result in a larger value than the observed upturn below 0.2 K. Moreover, decoupled Heisenberg antiferromagnetic chains would show a broad peak in magnetic susceptibility associated with the formation of magnetic correlations according to the Bonner-Fisher model [28], in contrast to the observed monotonic susceptibility down to 1.7 K, implying that the broad peak is suppressed by strong frustration. A satisfactory description of $Ba_4Ir_3O_{10}$ must therefore include inter-chain coupling through the trimer midpoint sites, including also via further-neighbor exchange paths, with a scale J' that must at least satisfy J' >> 0.2 K and also be not much smaller than intra-chain coupling. The low symmetry implies that the inter-chain coupling will include SOI exchanges [1, 3, 5-9], beyond Heisenberg. This SOI frustration can complement geometrical frustration from further-neighbor exchanges to produce inter-chain coupling that is highly frustrated.

We thus turn to the quantum liquid states that can be theoretically described by adding coupling (of arbitrary SOI forms) to an array of spins and Luttinger liquids. The first possibility is that even with moderate inter-chain couplings the effective physics is similar to that of decoupled chains. Such a phase is known as a *sliding Luttinger liquid* [29] and remains stable at least down to ultralow temperatures, where high order instabilities appear. (Note that with complicated inter-chain interactions a true sliding phase can occur where all instabilities are irrelevant [30]). Indeed, numerical studies of Heisenberg coupled spin chains find that even moderate inter-chain couplings ($J' = rJ$ with $r \approx 0.3 – 0.7$) can be renormalized by frustration [31] to preserve approximate 1D Luttinger liquid physics. The second possibility is that the additional trimer-midpoint spins and couplings drive the system into a 2D quantum spin liquid phase, such as a *spinon Fermi surface* [32]. Interestingly, the observed T-linear heat capacity is consistent with sliding Luttinger liquids but inconsistent with the $T^{2/3}$ leading order term of a spinon Fermi surface [33]. However, this apparent inconsistency can be resolved if the next order T-linear contributions become important or if crystalline disorder gives a sufficiently short spinon mean free path so that the corresponding energy scale is larger than 2.5 K, bounding the spinon conductivity and giving up to logarithmic corrections, T-linear heat capacity at lower temperatures. The observed T-linear thermal conductivity (**Fig.5b**) implies metallic behavior consistent with both possibilities. The constant T=0 offset and its extensive entropy implies, independently of any particular model, some collection of oscillator modes with ultralow frequencies $\omega < 0.2$ K such that already by T = 0.2 K these modes are thermally excited and produce constant C(T). The experimental data directly requires such an ultralow energy scale $\omega$; let us now consider how it can arise in the two possibilities of our theoretical model. A scenario consistent with both possibilities would be proximity to a quantum critical point with a collection of bosonic critical fluctuations. In a spinon Fermi surface, a band of bosonic excitations known as the emergent photon arises, and could potentially produce the C(T) offset if its velocity is small enough. In contrast, excitations in a sliding Luttinger liquid could have very different velocities, which could produce behavior similar to that observed. In addition to heat capacity, thermal transport could in principle distinguish the two possibilities, though extracting detailed constraints is complicated by uncertainty in the role of disorder. Future experiments should distinguish between the different scenarios for the quantum liquid, most directly by measuring the inelastic spin structure factor, which could distinguish between 2D fractionalized spinons and 2D descendants of 1D spinons.

The numerical magnitude of the linear heat capacity provides additional support for both quantum liquid scenarios. (In contrast, the magnitude of thermal conductivity involves non-universal scattering factors.) The slope of C(T) is $\gamma = 17$ mJ / mole K$^2$ = 1/(490 K) per formula unit. This already provides an energy scale. Numerical pre-factors are determined in each given theory, for example in Luttinger liquid theory this slope is associated with the Fermi velocity of the gapless fractionalized excitations:

$$C[T] = \frac{\pi N}{3} \frac{a}{\hbar v} k_B^2 T \quad (2)$$

with $a$ the lattice constant of the chains, $v$ the velocity of the excitations, and $N=2$ the number of relevant (trimer endpoint) Ir ions per formula unit, giving a velocity with a characteristic band-edge energy scale $\hbar v / k_B a = 1000$ K. These extracted energy scales are at the same scale as the Curie-Weiss exchange energy and also correspond to the scale of exchange energies expected for corner-sharing $IrO_6$ octahedra compounds such as $Sr_2IrO_4$, further supporting the picture of Luttinger liquid chains consisting of corner-sharing octahedra of trimer-endpoint spins.

This study uncovers an exotic quantum liquid state emerging in the magnetic insulator $Ba_4Ir_3O_{10}$. The state is hallmarked by T-linear heat capacity and unusual thermal conductivity with a lack of long-range order down to 0.2 K. Despite the enormous thousand-fold frustration ratio, the seemingly obvious Ir trimer units in this structure form a square lattice with little frustration. We propose a new frustration mechanism that leads to the quantum liquid state. A reduced intra-trimer exchange produces an array of coupled 1D Tomonaga-Luttinger liquids with additional spins, capturing most of the striking experimental results even in the decoupled limit and supporting the formation of a 2D quantum liquid state. Indeed, slight chemical doping is experimentally found to readily recover semiclassical magnetic order and destroy the signatures of the quantum liquid state. This discovery provides a much-needed new paradigm for the search of quantum liquids in insulators, opening the door to both new experimental avenues and to theoretical questions about newly relevant microscopic models and possible new phases of quantum matter.

**Methods.** Single crystals of $Ba_4Ir_3O_{10}$ (and Sr doped $Ba_4Ir_3O_{10}$) were grown using a flux method from off-stoichiometric quantities of $IrO_2$, $BaCO_3$ (and $SrCO_3$) and $BaCl_2$ in a weight ratio of 1:10:2.7. The mixture was fired at 1460 C for 10 hours and then slowly cooled to room temperature at a rate of 4 C/hour. Platinum crucibles were used. The average sample size is approximately 3 x 2 x 1 $mm^3$. Measurements of crystal structures were performed using a Bruker Quest ECO single-crystal diffractometer equipped with a PHOTON 50 CMOS detector. It is also equipped with an Oxford Cryosystem that creates sample temperature environments ranging from 80 K to 400 K during x-ray diffraction measurements. Chemical analyses of the samples were performed using a combination of a Hitachi MT3030 Plus Scanning Electron Microscope and an Oxford Energy Dispersive X-Ray Spectroscopy (EDX). Magnetic properties were measured using a Quantum Design (QD) MPMS-7 SQUID Magnetometer with a rotator that enables angular measurements of magnetic properties. Standard four-lead measurements of the electrical resistivity were carried out using a QD Dynacool PPMS System equipped with a 14-Tesla magnet. The heat capacity was measured down to 0.05 K using a dilution refrigerator for the PPMS.

Thermal conductivity measurements were performed with a one-heater, two-thermometer configuration, with the temperature gradient along the *c* axis. Cernox resistors were used as thermometers for the reported temperature range (T > 1.8 K) and calibrated separately in situ under the applied field. The sample was mounted in a mini-vacuum cell, which was specially designed to be mounted in the rotation probe, that enables the external magnetic field to be aligned parallel with the c-axis [i.e. along the temperature gradient] and perpendicular to the out-of-plane direction. Temperature gradient $|\Delta T|$ is maintained below 3% of the bath temperature to avoid heating.

The crystal structure of $Ba_4Ir_3O_{10}$ was also studied using single crystal neutron diffraction using the time-of-flight Laue diffractometer TOPAZ. The data were collected on crystal with a volume of approximately 1.0 mm$^3$ for 70 minutes at every orientation. A total of approximately 20 hours was spent to collect the data at 100 K. Sample orientations were optimized for an estimated 99% coverage of symmetry-independent reflections of the monoclinic cell. The raw Bragg intensities were obtained using a 3D ellipsoidal integration method. Data reduction including Lorentz, absorption, TOF spectrum, and detector efficiency corrections were carried out. The reduced data were exported to the GSAS program suite for wavelength dependent extinction correction and refined to convergence using SHELXL-97.

**Data Availability.** The data that support the findings of this study are available from the corresponding author upon reasonable request.

**Acknowledgements.** We acknowledge very useful discussions with Drs. Daniel Khomskii, Sergey Streltsove, Igor Mazin, Leon Balents and Bernd Buchner. The experimental work conducted in G.C.'s group was supported by NSF grants DMR-1712101 and 1903888. Thermal conductivity data were obtained in the National High Magnetic Field Laboratory, which is supported by NSF Cooperative Agreements No. DMR-1157490 and the State of Florida. M.H. was supported by the U.S. Department of Energy, Office of Science, Basic Energy Sciences (BES) under Award number DE-SC0014415. I.K. was supported as follows: until 12/10/2018, in part by a Simons Investigator Award to Leo Radzihovsky from the James Simons Foundation, and in part by the Army Research Office under Grant

Number W911NF-17-1-0482; after 12/10/2018, by a National Research Council Fellowship through the National Institute of Standards and Technology. The views and conclusions contained in this document are those of the authors and should not be interpreted as representing the official policies, either expressed or implied, of the Army Research Office or the U.S. Government. The U.S. Government is authorized to reproduce and distribute reprints for Government purposes notwithstanding any copyright notation herein. The work at ORNL was sponsored by the Scientific User Facilities Division, Office of Basic Energy Sciences, U.S. Department of Energy. C. A. P was partially supported by Colorado Energy Research Collaboratory.

**Competing Interests.** The authors declare no Competing Financial or Non-Financial Interests.

**Author Contributions.** G.C. grew the single crystals, performed measurements of physical properties on them and the analysis of the experimental data, directed the project and, together with I.K., wrote the paper with input from all authors. I.K. provided the theoretical model/analysis and, with G.C., wrote the paper with input from all authors. H.Z. performed measurements of heat capacity, resistivity and magnetization. H.D.Z. performed measurements of the crystal structure using single-crystal X-ray diffraction and heat capacity. F.Y. performed neutron diffraction of the crystals and provided the analysis of the crystal structure. C.A.P. performed measurements of thermal conductivity and M.Y.L. directed these measurements and provided the analysis of the thermal conductivity data and other data. M.H. provided the theoretical analysis and critical insights into the physics of the system. Y.F.N, Y.Z., C.H., and X.P.W. made equal contributions to measurements of the crystal structure.

**Figure Captions:**

**Fig.1. Crystal structure of $Ba_4Ir_3O_{10}$:** **(a)** *ab* plane. Ir sites with O octahedra (red/green) at top; Ir ions (blue) highlighted below, showing the wavy 2D structure. **(b)** *bc* plane. Colors as in (a), here showing the Ir trimers. The resulting lattice of trimers has no geometric frustration. **(c)** A representative single crystal of $Ba_4Ir_3O_{10}$.

**Fig.2. Magnetic susceptibility of $Ba_4Ir_3O_{10}$ (x = 0):** Temperature dependence of **(a)** magnetic susceptibility $\chi$ at $\mu_oH = 1$ T for the *a*, *b*, and *c* axes; Inset: same data in semi-log plot. **(b)** $1/\Delta\chi$, where $\Delta\chi = \chi - \chi_o$ with $\chi_o$ the temperature-independent (Van Vleck) susceptibility, for the *a*, *b*, and *c* axis for $100 \leq T \leq 350$ K. The Curie-Weiss temperatures $\theta_{CW}$ are -766 K, -390 K, and -169 K, indicating a strong AFM interaction and magnetic anisotropy, yet no sign of long-range order.

**Fig.3. Low-temperature heat capacity of $Ba_4Ir_3O_{10}$ (x = 0):** Temperature dependence of **(a)** heat capacity C(T) for $0.05$ K $\leq T \leq 4$ K at $\mu_oH = 0$ T; **(b)** C(T) at $\mu_oH = 0.2$ T applied along the *a*, *b*, and *c* axis, respectively. Inset: C at 0.052 K as a function of H. **(c)** C at $\mu_oH$ = 2 T or 3T aligned along the *a*, *b*, and *c*-axis, respectively; the yellow-green dashed line is C(T) at H = 0 for comparison. Observe that the pronounced linear dependence of C(T) (highlighted in (a) by the blue dashed line) spans a decade in temperature from 0.2 to 2.5 K with a sizable slope $\gamma$.

**Fig. 4. Magnetic properties of x = 0.02 and comparison with x = 0: (a)** The temperature dependence of the magnetization M for the *a, b* and *c* axes at $\mu_oH = 1$ T (left scale) and heat capacity C (right scale) for x = 0.02; Inset: the isothermal magnetization M at T = 1.8 K for the *a, b* and *c* axes for x = 0.02. Long-range AFM order occurs at $T_N$ = 130 K, with another anomaly at $T_M$. **(b)** Angular dependence of magnetization M for x = 0.01 and 0.02

at 6 K and 5 T. The sharp angular dependence clearly defines the easy axis along the $a$ axis.

**(c)** $1/\Delta\chi$ for comparison between x = 0 and x = 0.02. Note that the frustration parameter $f$ is reduced by three orders of magnitude upon Sr doping.

**Fig.5. Thermal properties of x = 0.01 and 0.02 and comparison with x = 0: (a)** C(T) for 0.05 K $\leq$ T $\leq$ 4 K for comparison between x = 0, 0.01 and 0.02. Note that the linearity that defines C(T) for x = 0 vanishes in x = 0.01 and 0.02 (Inset). **(b)** The $c$-axis thermal conductivity $k_c$(T) for x =0 and x = 0.02. Note that the drastic changes in $k_c$(T) upon Sr doping; in particular, the distinct low-temperature linearity in x = 0 and disappearance of it in x = 0.02 (Inset).

**Fig.6. Hopping, trigonal distortion and lattice recombination: (a)** Ir-O-Ir hopping amplitude $t$ for face-sharing octahedra (dotted purple curve) as a function of trigonally distorted Ir-O-Ir angle $\theta$, and corresponding Ir-Ir magnetic exchange $J'\sim t^2$ (solid blue curve). ($\theta \approx 70.5°$ for ideal octahedra; $\theta \approx 73° - 79°$ in $Ba_4Ir_3O_{10}$.) Hopping is computed in a tight-binding limit following Ref. [28] ($\frac{<r_{Ir}^2>}{<r_{Ir}^4>} = \frac{0.3}{a_B^2}$, $|r_{Ir-O}| = 2.1$Å) with a single hole in the low energy $e_g^\pi$ states; adding SOI and distortions beyond trigonal will project to a Kramer's doublet primarily involving $e_g^\pi$ and modify the plotted curve, likely shifting the sign change in $\theta$. The key qualitative feature is the hopping sign change and associated region of reduced magnetic exchange (green shaded oval), providing a mechanism for the lattice recombination shown in **(b)**. **(b)** The same view as Fig. 1b: trimer units (ovals on top) recombine into c-axis 1D zigzag chains (solid lines) that couple via the remaining trimer-midpoint spins (dashed lines).

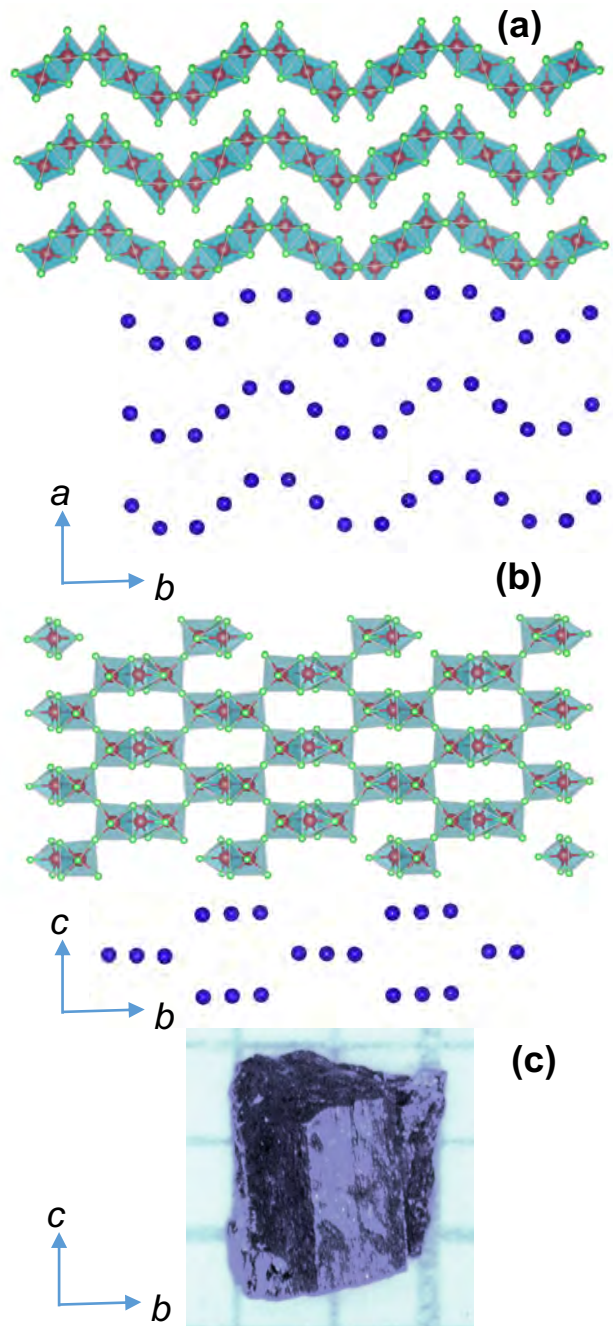

Fig.1

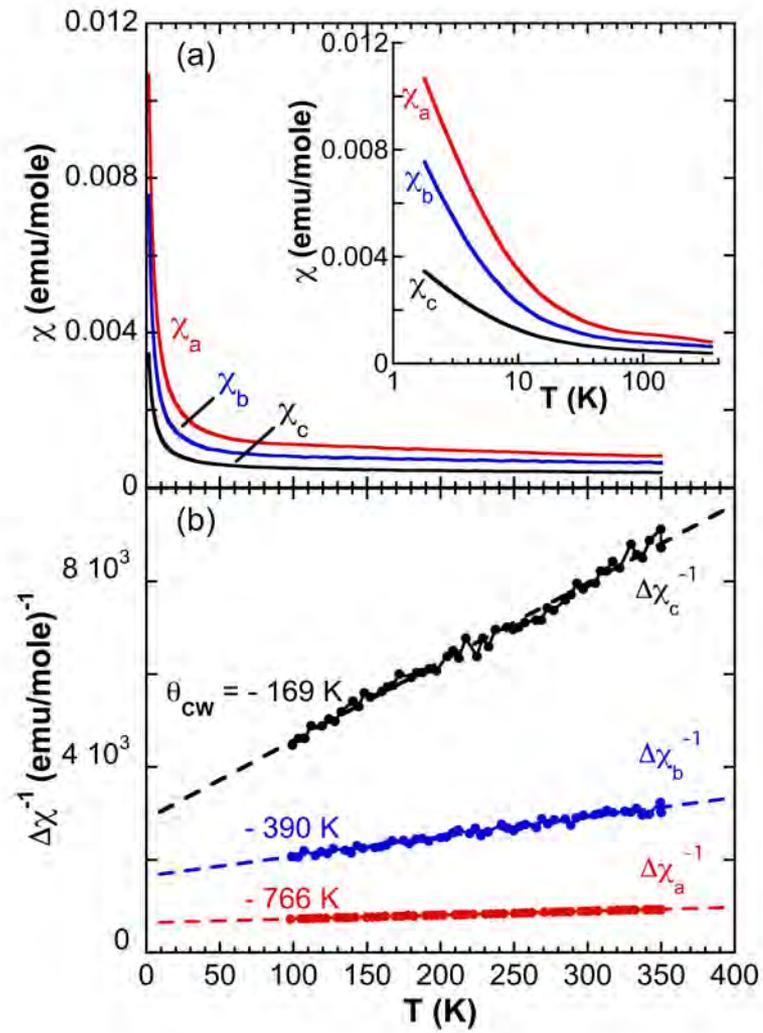

Fig.2

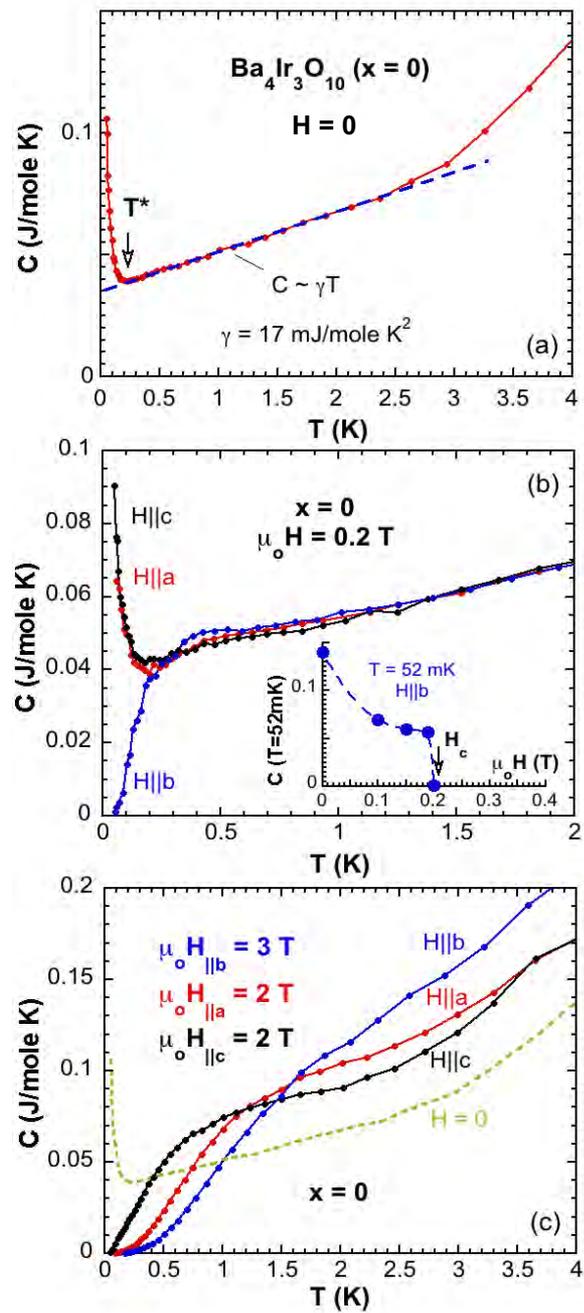

Fig.3

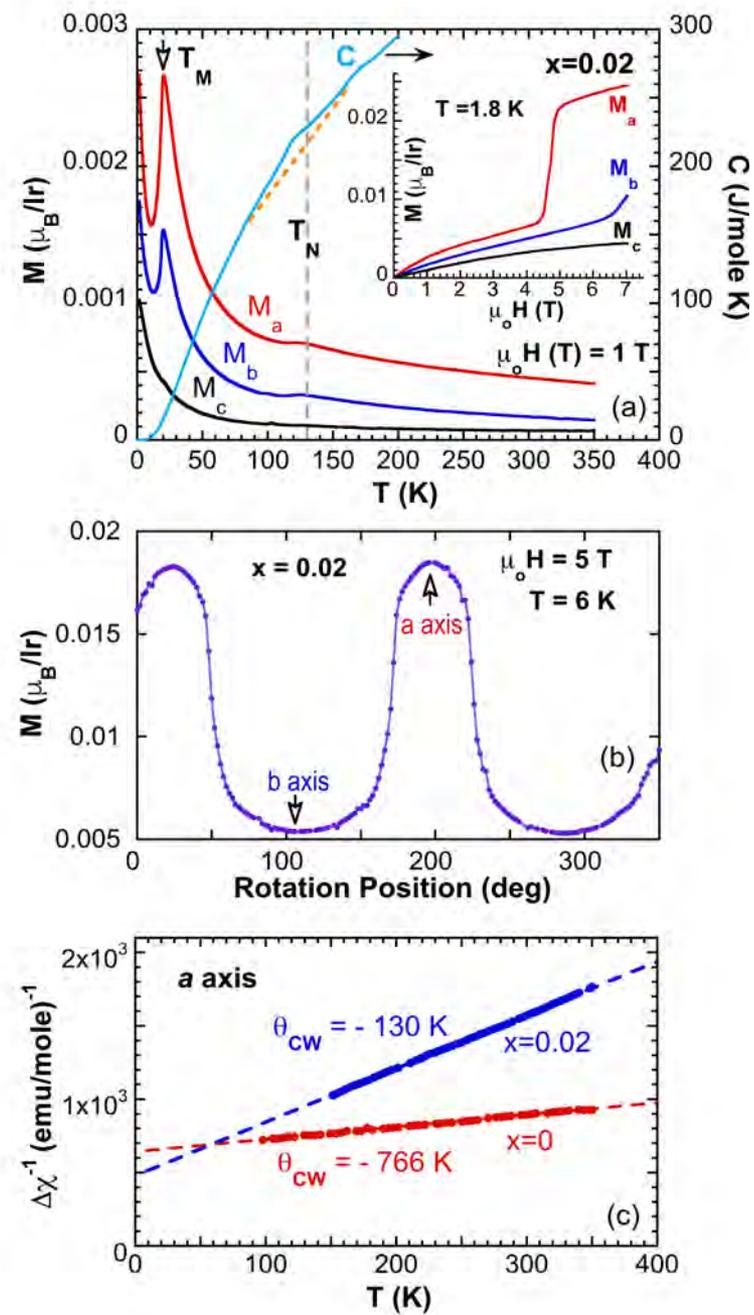

Fig.4

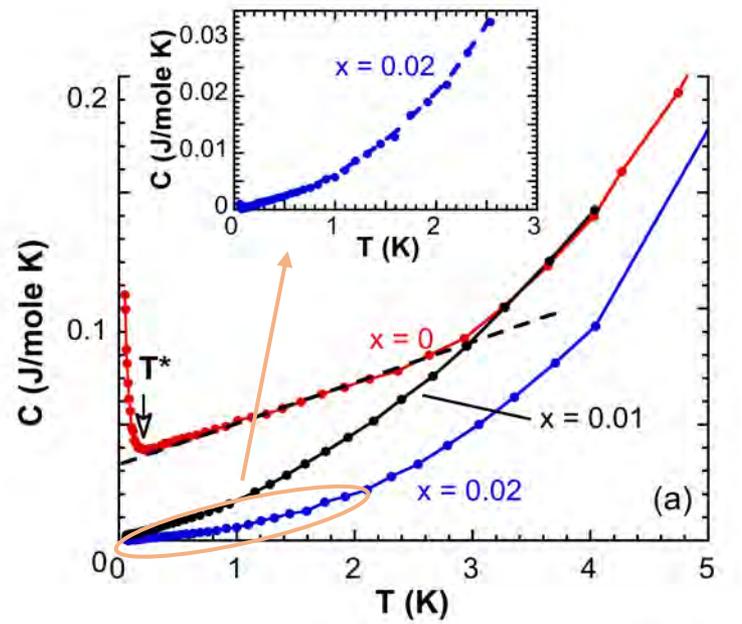
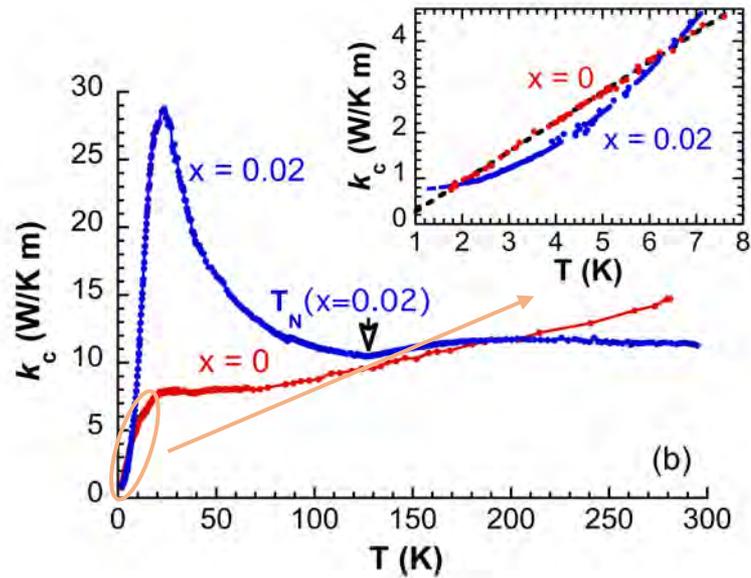

Fig.5

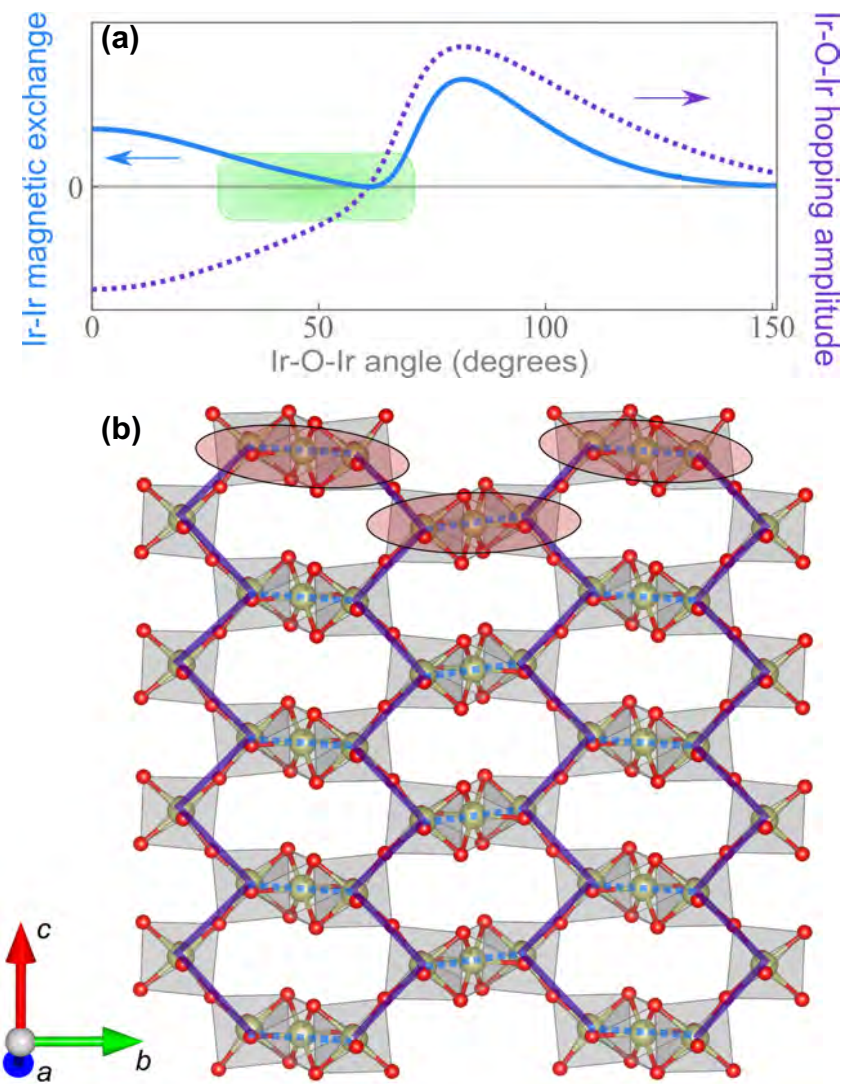

Fig.6